\documentclass[aps,pra,showpacs,twocolumn,superscriptaddress,floatfix,fleqn,byrevtex]{revtex4}
\usepackage{amsmath,amssymb}
\usepackage{graphics,color,epsfig}
\renewcommand{\Im}{\mathop{\rm Im}}
\renewcommand{\Re}{\mathop{\rm Re}}
\newcommand{\ka}{\hbox{\ae}}

\begin{document}
\title{Manipulative resonant nonlinear optics: nonlinear interference effects and quantum control of nonlinearity, dispersion,
transparency and inversionless amplification in an extended
strongly-absorbing inhomogeneously-broadened medium}
\author{A.~K.~Popov}\email{apopov@uwsp.edu}
\homepage{http://www.kirensky.ru/popov}\affiliation{Department of Physics \& Astronomy
and Department of Chemistry,
University of Wisconsin-Stevens Point, Stevens Point, WI 54481,
USA} \affiliation{Institute of
Physics of the Russian Academy of Sciences, 660036 Krasnoyarsk,
Russia}
\author{S.~A.~Myslivets}\affiliation{Institute of Physics
of the Russian Academy of Sciences, 660036 Krasnoyarsk, Russia}
\author{Thomas~ F.~ George}\email{tfgeorge@umsl.edu}
\homepage{http://www.umsl.edu/chancellor} \affiliation{Department
of Chemistry \& Biochemistry and Department of Physics \&
Astronomy, University of Missouri-St.~Louis, St.~Louis, MO 63121,
USA}
\date{July 29, 2004}
\begin{abstract}
Specific features of nonlinear interference processes at quantum
transitions in near- and fully-resonant optically-dense
Doppler-broadened medium are studied.  The feasibility of
overcoming of the fundamental limitation on a velocity-interval of
resonantly coupled molecules imposed by the Doppler effect is
shown based on quantum coherence. This increases the efficiency of
nonlinear-optical processes in atomic and molecular gases that
possess the most narrow and strongest resonances. The possibility
of all-optical switching of the medium to opaque or,
alternatively, to absolutely transparent, or even to
strongly-amplifying states is explored, which is controlled by a
small variation of two driving radiations. The required
intensities of the control fields are shown to be typical for cw
lasers. These effects are associated with four-wave mixing
accompanied by Stokes gain and by their interference in fully- and
near-resonant optically-dense far-from-degenerate double-Lambda
medium. Optimum conditions for inversionless amplification of
short-wavelength radiation above the oscillation threshold at the
expense of the longer-wavelength control fields, as well as for
Raman gain of the generated idle infrared radiation, are
investigated. The outcomes are illustrated with numerical
simulations applied to sodium dimer vapor. Similar schemes can be
realized in doped solids and in fiber optics.
\end{abstract}
\maketitle
\section{INTRODUCTION}
In the early stages of the laser era four decades ago, the major
problem was extracting coherent optical electromagnetic radiation
from incoherently-excited materials. The contemporary trend is the
inverse, i.e., control of physical properties of materials by
coherence injected with lasers. In such a context, this paper is
aimed at studying the opportunities of manipulating optical and
nonlinear-optical properties of materials without changing their
composition and structure based on quantum coherence and
interference. The phenomena of resonance, Doppler shift of a
frequency, coherence and interference are fundamental concepts of
physics. Resonance enables one to greatly enhance processes
related to oscillations. The strongest resonances in optics are
attributed to quantum transitions in free atoms and molecules.
However, one of the fundamental limitations of optical physics is
that only a small fraction of molecules can be resonantly coupled
with coherent radiation concurrently in an
inhomogeneously-broadened medium. This paper shows that coherent
control enables one to remove such limitations and thus
substantially increase the efficiency of the nonlinear optical
processes.

Coherence is an important concept in physics. Coherent
oscillations may interfere to give rise to counter-intuitive
effects. A common example is a standing light wave:
\begin{eqnarray*}
y(t)&=y_1(t)+y_2(t)
= a \sin(\omega t - kz) + a\sin(\omega t + kz)\\
&=2a\cos(\omega t)\cos(kz).
\end{eqnarray*}
By overlapping two coherent light fluxes, one can produce
completely-dark spots, instead of achieving doubled illumination,
or vice versa, one can achieve a twofold increase of the
oscillation amplitude, which in turn yields a fourfold increase in
the illumination. The concept of interference, which is a measure
of coherence, is applied to oscillations of any nature. Similar
effects can be anticipated in regard to oscillations of bound
intra-atomic and molecular charges. Since induced
periodically-accelerated and decelerated motion of the atomic
electrons is responsible for optical process, the question is how
to manage destructive interference and completely stop induced
oscillations of an optical electron even in the presence of a
resonant electromagnetic field. This would mean the decoupling of
light and resonant media, vanished absorption and dispersion.
Then, if destructive interference is turned into constructive,
that would provide for strongly-enhanced coupling.

The basic idea behind such opportunities can be presented as
follows. The oscillating dipole moment of the unity volume,
$P(t)$, is a driving term in Maxwell's electrodynamic equations
that determines the optical properties of the materials. It is
proportional to the number density of atoms $N$ and to the induced
atomic dipole moment $d(t)$. If, besides the probe radiation
$E_p(t)$, an auxiliary strong driving field $E_d(t)$ is turned on
[$E_{p,d}(t)=E_{p,d}$ $\exp(i\omega_{p,d}t)$], optical electrons
become involved in nonlinear oscillations:
\begin{equation}
P(t) = \chi E_p (t) + \chi^{(3)} |E_d(t)|^2 E_p(t).\label{os}
\end{equation}
This equation displays the superposition of two components
oscillating with the same frequency. In the vicinity of a
resonance, the real and imaginary parts of the susceptibilities
can be commensurate. Therefore, one can control the amplitudes and
relative phase of the interfering oscillations (i.e., constructive
and destructive interference) by varying the intensity and
detuning from the resonance of the driving field. In quantum
terms, this means that the coupling of light and matter and,
therefore, optical properties of the materials, can be controlled
through judicious manipulation of the interference of one-photon
and multiphoton quantum pathways. Consequently, this enables one
to  selectively deposit energy in certain quantum states or,
alternatively, to convert energy incoherently deposited in
specific quantum states into coherent light, while decoupling such
radiation from the absorbing atoms at resonant lower levels. As
such, spectral structures in the transparency and refractive index
can be formed and modified with the aid of the auxiliary control
resonant or quasi-resonant radiation.

Coherent quantum control of optical processes has proven to be a
powerful tool to manipulate refraction, absorption, transparency,
gain and conversion of electromagnetic radiation (for a review
see, e.g.,
\cite{{1Sc},{1Ko},{1Ma},{1Fl},{1RSPI},{1BRAN},{1PSPI},{1Har}}).
Among the recent achievements are the slowing down of the light
group speed to a few m/s, highly-efficient frequency
up-conversion, squeezed quantum state light sources and optical
switches for quantum information processing on this basis \cite
{{2Mer}, {2Hau}, {2Yam}, {2Kas}, {2Mat}, {Sw},{PMG}}, and
judicious control and feasibilities of near 100\% population
transfer between the quantum energy levels \cite{Berg}. Much
interest has been shown in the physics and diverse practical
schemes of lasing without population inversion (see,
e.g.,\cite{{1Sc},{1Ko},{1Ma},{1Fl},{1RSPI},{1BRAN},{1PSPI},{Corb}}).
Large enhancement of fully-resonant four-wave mixing through
quantum control via continuum states has been proposed in
\cite{PKG04}.

The present paper considers optical processes controlled through
"injected" quantum coherence, which is followed by destructive and
constructive nonlinear interference effects. The near- and
fully-resonant far-from-degenerate double-Lambda  scheme is
investigated, where inhomogeneous broadening of all coupled
(including multiphoton) transitions is accounted for. Analytical
solutions for density-matrix elements (level populations and
coherences) are found that account for various relaxation
channels, incoherent excitation of the coupled levels, and Doppler
shifts of the resonances. These shifts depend on the relative
direction of propagation of the coupled waves. The solution is
convenient for further numerical analysis and experiments. It is
shown that inhomogeneous broadening of Raman transitions, which is
inherent for far-from-degenerate schemes, creates important
features on quantum interference in resonant schemes. The
feasibility of overcoming with the aid of the coherence processes
of the seemingly fundamental limitation imposed by the Doppler
effect on the velocity-interval of the molecules concurrently
involved in resonant nonlinear coupling is shown. Thus, a major
part of Maxwell's velocity distribution can be involved in the
concurrent interaction that dramatically increases the
cross-section of nonlinear-optical processes in warm gases. As an
application, the conversion of the easily-achievable
long-wavelength gain to a higher frequency interval is explored.
It is shown that the most favorable are the optically-dense
samples, where the inhomogeneity of the field intensities along
the medium imposes other important features that must be accounted
for as well. The optimum conditions for the realization of such
processes are investigated, and the potential of quantum
switching, amplification and lasing without population inversion
based on coherent coupling at inhomogeneously-broadened
transitions in optically-dense spatially inhomogeneous media is
demonstrated with the aid of numerical simulations. Another
important outcome investigated in this paper is the feasibility of
manipulation of the resonant medium, from becoming almost entirely
optically opaque to strongly amplifying via a completely
transparent state by only a small variation of the intensity or
frequency of one of the lower-frequency control fields and of a
large gain for an idler infrared radiation. It is shown that the
required conditions can be fulfilled, e.g., in the framework of
experiments similar to \cite{3Hin} with sodium dimer vapor. The
important features of both the signal and idler radiation inherent
for such a scheme include entangled states and suppression of
quantum noise.

This paper is organized into seven major sections. Section
\ref{pexp} describes a principal experimental scheme aimed at
manipulating optical transparency of initially strongly-absorbing
media, and provides a set of equations for coupled electromagnetic
waves and their solution for several limiting cases. A
corresponding theoretical model and the underlying physics are
discussed. Section \ref{dm} gives the density matrix equations.
Their solutions related to the problems under consideration are
given in Appendix A. The physical principles of manipulating the
local absorption (amplification) and refractive indices, as well
as the principles of eliminating inhomogeneous broadening of
nonlinear resonances, are described in Sec. \ref{ncie}. Results of
numerical experiments towards the outlined effects are presented
in Sec. \ref{ns}. The emphasis is placed on optical switching and
amplification without population inversion in an optically-dense
medium assisted by Stokes gain, as well as on the optimal
conditions to achieve the effects. Section \ref{rcqc} outlines
earlier experiments on quantum control based on constructive and
destructive interference and is to show a great variety of
applications of similar schemes of quantum control. These include
up-conversion of infrared, and generation of short-wavelength VUV
radiation, selective detection of relaxation processes and weak
electric and magnetic dc fields, as well as selective population
of energy levels. Section \ref{c} summarizes the main outcomes of
the paper.
\section{Resonant four-wave mixing assisted by Stokes gain}\label{pexp}
Consider a simple experiment designed to explore the manipulation
of optical properties through quantum interference based on the
experimental setup described in a number of publications on
resonant nonlinear optics in sodium vapor (e.g. \cite{3Hin}).
Two-atom sodium molecules (dimers), Na$_2$, possess an
electronic-vibration-rotation spectrum in the near-IR and visible
ranges that can be covered with commercial frequency-tunable cw
lasers.
\begin{figure}[!h]
\includegraphics[height=.3\textwidth]{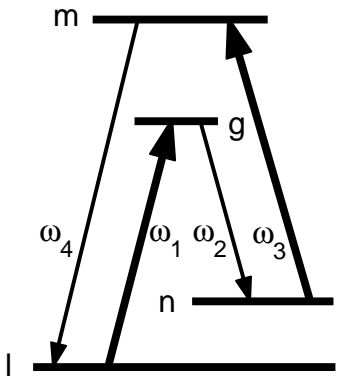}
\caption{\label{f1}Optical switching at $\omega_4$ through quantum
control by two fields at $\omega_1$ and $\omega_3$ assisted by
idler radiation at $\omega_2$.}
\end{figure}
Figure \ref{f1} depicts a relevant partial scheme of energy levels
and transitions. Similar schemes are typical for solids doped by
rare earths (like Er$^{3+}$ and Pr$^{3+}$). Assume that the sample
is probed with a sufficiently-weak (and thus non-perturbative)
optical field in the vicinity of the blue transition $ml$. Then
the refractive and absorption indices in the corresponding
frequency interval can be controlled with the driving fields
coupled to the lower or upper levels. Resonant driving fields
induce back-and-forth transitions with the frequency
$G_i=E_id_{jk}/2\hbar$ between levels $j$ and $k$, where $G_i$
(called the Rabi frequency) is the magnitude of the corresponding
interaction Hamiltonian scaled to Planck's constant. Modulation of
the probability amplitudes leads to a splitting of the resonance
at the probe transition and to interference of the one- and
two-photon quantum pathways controlled by the driving radiation.
This may exhibit itself as odd coherence and interference
structures in the absorption and refraction spectra of the probe
radiation.

Two independently-varied fields provide for even more
opportunities for manipulating local optical parameters. But three
coupled fields, a probe and two driving fields, may generate
radiation in the vicinity of the transition $gn$ through four-wave
mixing (FWM) processes $\omega_4 + \omega_2 = \omega_1 +
\omega_3$. This idler radiation may actually not get depleted but
rather substantially enhanced, since the driving field $E_1$ can
easily produce a large Stokes gain. This gain, in turn, can be
controlled with the driving field $E_3$. These processes
dramatically change the propagation properties of the probe field,
because three fields,  enhanced Stokes radiation and two driving
fields at $\omega_2$, $\omega_1$ and $\omega_3$, contribute back
to the probe field through the FWM process. Thus, doubled FWM
assisted by Stokes amplification processes opens extra channels
for the transfer of energy from the driving fields to the probe
one and, consequently, allows control of the transparency of the
medium for the probe radiation. Such a process is usually called
optical parametric amplification (OPA).  However, as shown below,
in resonance, one cannot think about the entire process as a
sequence of conventional Raman and FWM processes. This would be
misleading in predicting and interpreting the relevant
experiments. Indeed, the distinguishing differences in the
frequency-correlation properties of single- and multiphoton
radiation processes dramatically change in strong resonant fields.
This leads to a variety of important effects and applications that
are discussed below. Thus, the the effect under consideration
should be viewed as amplification without inversion (AWI) rather
than conventional OPA accompanied by absorption and Stokes gain.
\textit{The quantum interference involved in resonant schemes
plays such a crucial role that  thinking about the conversion
process under investigation in terms of the Manley-Rowe photon
conservation law would be misleading} \cite{5Tar}.

The propagation of four coupled optical waves in a resonant medium
is described by the set of coupled equations for the waves
$(E_i/2)\, \exp [i(k_iz-\omega_i t)] + c. c.\, \, (i= 1...4)$
traveling in an optically-thick medium:
\begin{align}
&d E_{4,2}(z)/dz=i\sigma_{4,2} E_{4,2}+i\tilde\sigma_{4,2}
E_1E_3E_{2,4}^*, \label{e42}\\
&d E_{1,3}(z)/{dz}=i\sigma_{1,3} E_{1,3}+
i\tilde\sigma_{1,3}E_4E_2E_{3,1}^*. \label{e13}
\end{align}
Here, $\omega_4+\omega_2 =\omega_1+\omega_3$, $k_j$  are wave
numbers in a vacuum, $\sigma_j=-2\pi k_j \chi_j = \delta
k_j+i{\alpha_j}/{2}$, where $\alpha_j$ and $\delta k_j$ are the
intensity-dependent absorption indices and dispersion parts of the
wave numbers, $\tilde\sigma_4=-2\pi k_4 \tilde\chi_4 $ (etc.) is a
FWM cross-coupling parameter,  and $\chi_j$ and $\tilde\chi_j$ are
the corresponding nonlinear susceptibilities dressed by the
driving fields. Amplification or  absorption of any of the coupled
radiations influences the propagation features of the others. In
the following sections, we will account for possible depletion of
the fundamental beams along the medium due to absorption. However,
qualitative feature of amplification of the probe beam and
generation of the idler radiation are seen within the
approximation that a change of the driving radiations $ E _ {1,3}
$ along the medium is neglected. This is possible in some cases,
e.g., at the expense of the saturation effects. Then  the system
(\ref {e42})-(\ref {e13}) reduces to two coupled standard
equations of nonlinear optics for $ E _ {4} $ and $ E _ {2} $,
whereas the medium parameters are homogeneous along $ z $. The
solution is standard:
\begin{widetext}
\begin{eqnarray}
E_2^*=\exp\left(-\frac{\alpha_2}{2}\,z-\beta z\right)
\left\{E_{20}^*\left[ \cosh(Rz)+\frac{\beta}{R}
\sinh(Rz)\right]-i\frac{\gamma_2^*}{R} E_{40}\sinh(Rz)\right\},\nonumber\\
E_4=\exp\left(-\frac{\alpha_4}{2}\,z+\beta z\right)
\left\{E_{40}\left[\cosh(Rz)-\frac{\beta}{R}
\sinh(Rz)\right]+i\frac{\gamma_4}{R}
E_{20}^*\sinh(Rz)\right\}\label{smopa}.
\end{eqnarray}
\end{widetext}
Here, $R=\sqrt{\beta^2+\gamma^2}$, $\beta =
\left[(\alpha_4-\alpha_2)/2+i\Delta k\right]/2$, $\Delta k=\delta
k_1+\delta k_3-\delta k_2-\delta k_4$,
$\gamma^2=\gamma_2^*\gamma_4$, $\gamma_{4, 2}=\chi_{4, 2}E_1E_3$,
and $E_{20}^*$ and $E_{40}$ are input values (at $z=0$). The first
terms in the curved brackets indicate OPA, and the second terms
the processes of FWM.

If $\Delta k=\delta k_1+\delta k_3-\delta k_2-\delta k_4=0$, the
input values $I_{4,2}\equiv |E_{4,2}|^2$ at $z=0$ are $I_{20}$=0
and $I_{40}\neq 0$, and the absorption (gain) rate substantially
exceeds that of the nonlinear optical conversion obtained at the
exit of the medium of length $L$:
\begin{eqnarray}
I_4/I_{40}&=&\left|\exp(-\alpha_4l/2)+({\gamma^2}/{(2\beta)^2})\right.\nonumber\\
&\times&\left.\left[\exp(g_2L/2)-\exp(-\alpha_4L/2)\right]\right|^2.\label{opa}
\end{eqnarray}
Alternatively, if $I_{40}=0, I_{20}\neq 0$,
\begin{eqnarray}
\eta_4 =I_4/I_{20}&=&({|\gamma_4|^2}/{(2\beta)^2})\nonumber\\
&\times& \left|\exp(g_2L/2)-\exp(-\alpha_4 L/2)\right|^2.
\label{fwm}
\end{eqnarray}
Here, $\gamma^2=\gamma_2^*\gamma_4$,\
$\gamma_{4,2}=\chi_{4,2}E_1E_3$, $\beta = (\alpha_4-\alpha_2)/4$,
$(|\gamma^2|/\beta^2\ll 1)$, and $g_2\equiv -\alpha_2 $. From
(\ref {opa}) and (\ref {fwm}), it follows that at relatively small
lengths, the FWM coupling may even increase the depletion rate of
the probe radiation, depending on the signs of $ \Im \gamma _
{4,2} $ and $ \Re\gamma _ {4,2}$. In order to achieve
amplification, large optical lengths $L$ and significant Stokes
gain on the transition $gn$ ($ \exp (g _ 2L/2) \gg | {(2\beta) ^
2} / {\gamma ^ 2} | $), as well as effective FWM both at
$\omega_2$ and $\omega_4$, are required. As seen from
(\ref{smopa}),(\ref{opa}) and (\ref{fwm}), the evolution of the
coupled fields along the medium strongly depends on whether the
Stokes field is injected at the entrance to the medium, and if so,
on the value of the ratio of the intensities of the input probe
and Stokes fields.

\section{Density matrix equation and macroscopic parameters of the medium}\label{dm}
The macroscopic parameters in Eqs. (\ref{smopa}) and (\ref{opa})
are convenient for calculations with the density matrix technique,
which allows one to account for various relaxation and incoherent
excitation processes. The power-dependent susceptibility
$\chi_{4}$, responsible for absorption and refraction, can be
found as
\begin{equation}
P(\omega _{4})= \chi_{4}E_{4},\, P(\omega _{4})=N\,\rho
_{lm}\,{d}_{ml}+{ c.c.},
\end{equation}
where $P$ is the polarization of the medium oscillating with the
frequency $\omega _{4}$, $N$ is the number density of molecules,
${d}_{ml}$ is the electric dipole moment of the transition, and
$\rho _{lm}$ is the density matrix element. Other polarizations
are determined in the same way. The density matrix equations for a
mixture of pure quantum mechanical ensembles in the interaction
representation can be written in a general form as
\begin{eqnarray}\label{ro}
&L_{nn}\rho_{nn}=q_n-i[V, \rho]_{nn}+ \gamma_{mn}\rho_{mm},&\nonumber\\
&L_{lm}\rho_{lm}=L_4\rho_4=-i[V,\rho]_{lm}, L_{ij}=d/dt\,+\Gamma_{ij},&\\
& V_{lm}=G_{lm}\cdot\exp\{i[\Omega_4t-kz]\},\quad G_{lm}=-{\bf
E_4\cdot d_{lm}}/2\hbar,&\nonumber
\end{eqnarray}
where $ \Omega _ 4 = \omega _ 4,-\omega_ {ml} $ is the frequency
detuning from the corresponding resonance; $ \Gamma _ {mn} $ -
homogeneous half-widths of the corresponding transition (in the
collisionless regime $ \Gamma _ {mn} = (\Gamma_m + \Gamma_n)/2$);
$ \Gamma _ n
 = \sum _ j\gamma _ {nj} $ - inverse lifetimes of levels; $ \gamma _ {mn} $ -
rate of relaxation from level $m$ to $n$; and $ q _ n = \sum _ j w
_ {nj} r _ j $ - rate of incoherent excitation to state $ n $ from
underlying levels. The equations for the other elements are
written in the same way.

It is necessary to distinguish the open and closed energy-level
configurations. In the open case (where the lowest level is not
the ground state), the rate of incoherent excitation to various
levels by an external source essentially does not depend on the
rate of induced transitions between the considered levels.  In the
closed case (where the lowest level is the ground state), the
excitation rate to different levels and velocities depends on the
value and velocity distribution at other levels, which are
dependent on the intensity of the driving fields. For open
configurations, $ q _ i $ are primarily determined by the
population of the ground state and do not depend on the driving
fields. The Doppler shifts of the resonances ${\bf k}_i{\bf v}$
can be accounted for in the final formulas by substituting $\Omega
_ i$ for $\Omega _ i'=\Omega _ i-{\bf k}_i{\bf v}$ (${\bf v}$ is
the atomic velocity).

In a steady-state regime, the solution of a set of density-matrix
equations can be cast in the following form:
\begin{align*}
&\rho_{ii}=r_i,\, \rho_{lg}=r_1\cdot \exp(i\Omega_1t),\
\rho_{nm}=r_3\cdot
\exp(i\Omega_3t),\\
&\rho_{ng}=r_2\cdot
\exp(i\Omega_1t)+\tilde{r}_2\cdot \exp[i(\Omega_1+\Omega_3-\Omega_4)t],\\
&\rho_{lm}=r_4\cdot \exp(i\Omega_4t)+\tilde{r}_4\cdot
\exp[i(\Omega_1-\Omega_2+\Omega_3)t],\\
&\rho_{ln}=r_{12}\cdot \exp[i(\Omega_1-\Omega_2)t]+r_{43}\cdot
\exp[i(\Omega_4-\Omega_3)t].
\end{align*}
The density matrix amplitudes $r_i$ determine the absorption/gain
and refraction indexes, and $\tilde{r}_i$ determine the four-wave
mixing driving nonlinear polarizations. Then the problem reduces
to the set of algebraic equations
\begin{align}\label{od}
&P_2r_2=iG_2\Delta r_2-iG_3r_{32}^*+ir_{12}^*G_1,\nonumber\\
&d_2\tilde{r}_2=-iG_3r^*_{41}+ir^*_{43}G_1,\nonumber\\
&P_4r_4=i\left[G_4\Delta r_4-G_1r_{41}+r_{43}G_3\right],\nonumber\\
&d_4\tilde{r}_4=-iG_1r_{32}+ir_{12}G_3\,\nonumber\\
&P_{41}r_{41}=-iG_1^* r_4+ir_1^*G_4,\nonumber\\
&P_{43}r_{43}=-iG_4 r_3^*+ir_4G_3^*,\nonumber\\
&P_{32}r_{32}=-iG^*_2 r_3+ir_2^*G_3,\nonumber\\
&P_{12}r_{12}=-iG_1 r_2^*+ir_1G_2^*,\nonumber\\
&P_1r_1=iG_1\Delta r_1,\,P_3r_3= iG_3\Delta r_3,
\end{align}
\begin{align}\label{d}
&\Gamma_m r_m=-2\Re\{iG_3^*r_3\}+q_m ,\nonumber\\
&\Gamma_n r_n=-2\Re\{iG_3^*r_3\}+\gamma_{gn}r_g+\gamma_{mn}r_m+q_n,\nonumber\\
&\Gamma_g r_g=-2\Re\{iG_1^*r_1\}+q_g ,\nonumber\\
&\Gamma_l
r_l=-2\Re\{iG_1^*r_1\}+\gamma_{gl}r_g+\gamma_{ml}r_m+q_l,
\end{align}
where  $G_1=-{\bf E_1}{\bf d}_{lg}/2\hbar$,\, $G_2=-{\bf E_2}{\bf
d}_{gn}/2\hbar$,\, $G_3=-{\bf E_3}{\bf d}_{nm}/2\hbar$,\,
$G_4=-{\bf E_4}{\bf d}_{ml}/2\hbar$,
$P_1=\Gamma_{lg}+i\Omega_1$,\, $P_2=\Gamma_{ng}+i\Omega_2$,\,
$P_3=\Gamma_{nm}+i\Omega_3$,\, $P_4=\Gamma_{lm}+i\Omega_4$,\,
$P_{12}=\Gamma_{ln}+i(\Omega_1-\Omega_2)$,\,
$P_{43}=\Gamma_{ln}+i(\Omega_4-\Omega_3)$,\,
$P_{32}=\Gamma_{gm}+i(\Omega_3-\Omega_2)$,\,
$P_{41}=\Gamma_{gm}+i(\Omega_4-\Omega_1)$,
$d_2=\Gamma_{ng}+i(\Omega_1+\Omega_3-\Omega_4)$,\,
$d_4=\Gamma_{lm}+i(\Omega_1-\Omega_2+\Omega_3),$
$\Omega_1=\omega_1-\omega_{lg},\, \Omega_3=\omega_3-\omega_{mn},\,
\Omega_2=\omega_2-\omega_{gn},\, \Omega_4=\omega_4-\omega_{ml}$,
$\Delta r_1=r_l-r_g,\, \Delta r_2=r_n-r_g,\, \Delta r_3=r_n-r_m,\,
\Delta r_4=r_l-r_m$.

For a closed scheme, Eq. (\ref{d}), $r_l$ must be replaced by

\begin{equation}
r_l=1-r_n-r_g-r_m. \label{cl}
\end{equation}
The solution of these equations as applied to the problem under
consideration is given in Appendix A.
\section{Nonlinear interference effects in absorption and
amplification spectra}\label{ncie}
\subsection{Nonlinear interference effects in three-level schemes}\label{nief} Consider two
simple subcases, $G_3=0$ ($V$-scheme) and $G_1=0$
($\Lambda$-scheme). From (\ref{od}), it follows that
\begin{eqnarray}
&P_4r_4=i\left[G_4\Delta r_4-G_1r_{41}\right],&\nonumber\\
&P_{41}r_{41}=-i\left[G_1^*r_4-r_1^*G_4\right],&\label{v}
\end{eqnarray}
\begin{eqnarray}
&P_2r_2=i\left[G_2\Delta r_2+r_{12}^*G_1\right],&\nonumber\\
& P_{12}r_{12}=-i\left[G_1 r_2^*-r_1G_2^*\right].&\label{l}
\end{eqnarray}
The structure of (\ref{v}) and (\ref{l}) is the same as that of
(\ref{os}): they present the interference of two oscillations. At
$\Gamma_{gm}\rightarrow \infty$ and $\Gamma_{ln}\rightarrow
\infty$, the interference disappears. Consequently, the effect of
the driving fields reduces to a change in the level populations
(optical pumping). Indeed, the two-photon coherences $r_{41}$ and
$r_{12}$ are the source of quantum nonlinear interference effects
(NIE), which lead to AWI, Autler-Townes splitting of the
resonances and to related effects of electromagnetically-induced
transparency, coherent population trapping, and to corresponding
changes in the refraction index. From (\ref{v}) or (\ref{ch1}) at
$G_3=0$, it follows that
\begin{align}
\dfrac{\chi_{4}}{\chi_{4}^0}&=\dfrac{\Gamma_4}{P_4}\cdot
\dfrac{\Delta
r_4-g_1\Delta r_1}{\Delta n_4( 1+g_4)}\nonumber\\
&= \dfrac{\Gamma_4}{\Delta n_4}\cdot\dfrac {\Delta r_4
P_{41}-(\Delta r_1|G_1|^2/P_1)}{P_{41}P_4
+|G_1|^2},\label{nie}\\
\dfrac{\chi_{2}}{\chi_{2}^0}&=\dfrac{\Gamma_2}{P_2}\cdot
\dfrac{\Delta
r_2-g_3\Delta r_1}{\Delta n_2(1+g_2)}\nonumber\\
&= \dfrac{\Gamma_2}{\Delta n_2}\cdot\dfrac {\Delta r_2
P^*_{12}-(\Delta r_1|G_1|^2/P^*_1)}{P^*_{12}P_2
+|G_1|^2}.\label{niel}
\end{align}
Both equations are of similar structures, and we will briefly
review major NIE using the example of Eq. (\ref{niel}). The
effects of the driving field $G_1$ contributing to the change of
absorption and refraction indices can be classified as: (i) a
change of the populations $\Delta r_4(|G_1|^2)$, dominant at
$\Gamma_{41}\Gamma_{gm}\rightarrow \infty$, and (ii) a splitting
of the resonance (see the denominator in (\ref{nie}), dominant at
$\Delta r_1=0$. With an increase in the detuning $\Omega_1$, one
of the components of Autler-Townes splitting determines the
ac-Stark shift of the resonance.

As for any interference, NIE cause only redistribution absorption
and amplification over the frequency interval (line shape) but do
not change the frequency-integrated absorption (amplification).
The latter is determined by the population change only: $\int
({\chi_{4}}/{\chi_{4}^0})d\Omega_4=\pi{\Gamma_4}\,{\Delta
r_4}/{\Delta n_4}$. The term in (\ref{nie}) associated with
$r_m=n_m$ describes the line shape (probability) of pure emission,
where all the rest (at $r_m=0$) describe pure absorption. Indeed,
as was first emphasized in \cite{{Not},{Sok}}, this difference
determines NIE in absorption, emission and amplification spectra
which enable the \emph{entire elimination of absorption and
appearance of transparency at unequal populations of the levels at
the probed transition and to amplification without population
inversion in some frequency interval(s) at the expense of enhanced
absorption in other intervals}. As seen from (\ref{nie}), at
$\Omega_4=\Omega_1=0$, the corresponding requirements to achieve
transparency and then amplification without a change of sign of
$\Delta r_4$ are
\begin{equation}
\Delta r_1|G_1|^2/\Gamma_{gm}\Gamma_{lg}\ge\Delta r_4.\label{awi}
\end{equation}
Therefore, owing to NIE, population inversion between the initial
and final bare states is not required in order to achieve  AWI in
this case. The requirements for transparency at unequal
populations of levels at the probed resonant transition and that
for amplification without population inversion for Ne transitions
was investigated in \cite{AwiP} (see also \cite{{Vved}}) during
the course of early studies of NIE with He-Ne lasers. Then lasing
without population inversion was experimentally realized in
\cite{AwiB} in gas discharge at the transition $2s_2$--$2p_4$
($\lambda$ = 1.15 $\mu$) of Ne in the presence of the driving
field of another He-Ne laser resonant to the adjacent transition
$2s_2$--$2p_1$ ($\lambda$ = 1.52 $\mu$).

\textit{It is misleading to think about resonant processes in
$\Lambda$ and $V$ schemes as conventional Raman processes}. The
amplification and absorption acquire the features of two-photon
processes, while the driving field is detuned from resonance and
interference between one- and two-photon pathways vanishes.
Indeed, at $|\Omega_1|\approx|\Omega_4|\gg\Gamma_1,\Gamma_4$,
$|g_4|\ll1, |g_1|\ll1$, $P_4\approx i\Omega_4, P_1\approx
i\Omega_1\approx i\Omega_4$,  one obtains from (\ref{nie}) the
equation for the absorption (amplification) index $\alpha
(\Omega_4)$  reduced by its value $\alpha^{0}(0)$ at resonance at
$G_1=0$:
\begin{align}
&\dfrac{\alpha (\Omega_4)}{\alpha^{0}(0)}\approx
\dfrac{\Gamma_4^2\Delta r_4}{\Omega_4^2\Delta n_4} -
\Re\left\{\dfrac{\Gamma_4(\Delta r_4g_4+\Delta r_1 g_1)}
{i\Omega_4\Delta n_4 }\right\}\nonumber\\
&\approx \dfrac{\Gamma_4^2\Delta r_4}{\Omega_4^2\Delta
n_4}-\dfrac{\Gamma_4\Gamma_{14}}
{\Gamma_{14}^2+(\Omega_4-\Omega_1)^2}
\cdot\dfrac{|G_1|^2(\Delta r_1-\Delta r_4)}{\Omega_4^2\Delta n_4}\nonumber\\
&= \dfrac{\Gamma_{lm}^2( r_l-r_m)}{(n_l-n_m)\Omega_4^2}\nonumber\\
&-
\dfrac{\Gamma_{gm}\Gamma_{lm}}{\Gamma_{gm}^2+(\Omega_4-\Omega_1)^2}
\cdot\dfrac{|G_1|^2 ( r_m- r_g)}{\Omega_4^2 (n_l-n_m)}.
\label{ram}
\end{align}
The last terms in (\ref{ram}) describe Raman-like coupling and
originate both from the nominator and denominator in Eq.
(\ref{nie}). In this case, population inversion between the
initial and final bare states ($r_m=n_m>r_g$) is  required for
amplification of the probe field.

In more detail, the nature of NIE in the context of interference
of quantum pathways and modification of the frequency-correlation
properties of multiphoton processes in strong resonant fields was
discussed in \cite{{Vved},{Feok},{Rau}}. Numerical examples of
AWI, including effects related to the growth of the amplified
field, are given in \cite{Kuch} both for open and closed energy
level configurations.
\subsection{Quantum coherence and elimination of inhomogeneous broadening of
multiphoton transitions with light shifts}\label{edb} As outlined
above, two-photon coherence $r_{ln}$ plays a key role in the
processes under investigation. The important feature of the
far-from-frequency-degenerate interaction is that the
inhomogeneous (in our case, Doppler) broadening of a two-photon
transition is much greater than its homogeneous width,
$\zeta=\Gamma_{ln}/(\Delta^D_{lg}-\Delta^D_{ng})\ll 1$, where
$\Delta^D_{lg}$ and $\Delta^D_{ng}$ are the Doppler HWHM of the
corresponding transition. Therefore, only a $\zeta$ fraction of
the molecules are resonantly involved in the process.
Correspondingly, the magnitude and sign of the multiphoton
resonance detunings and, consequently, of the amplitude and phase
of the lower-state coherence $\rho_{nl}$, differ for molecules at
different velocities due to the Doppler shifts. This is not the
case in near-degenerate schemes. The interference of elementary
quantum pathways, accounting for Maxwell's velocity distribution
and saturation effects, results in a nontrivial dependence of the
macroscopic parameters on the intensities of the driving fields
and on the frequency detunings from the centers of the
inhomogeneously-broadened resonances. However, this limitation
that seems fundamental can be overcome by a judicious compensation
of Doppler shifts with ac-Stark shifts.

Consider the principles of such compensation for the example of
the absorption (gain) index $\alpha_2$.  In order to account for
Doppler effects, we must substitute all $\Omega_j$ in the above
formulas for $\Omega_j'=\Omega_j-\bf{k}_j\bf{v}$, where $\bf v$ is
the molecular velocity, and then perform velocity integration over
a Maxwell distribution. Corresponding factors $P_{j}$ must be
substituted for velocity-dependent factors
$P'_{j}=P_j(\Omega_j')$. Let us assume all detunings $\Omega_j$,
except $\Omega_1-\Omega_2$, to be much greater than the
corresponding transition Doppler widths. Then all factors $P'_{j}$
and $P'_{ij}$, except $P'_{12}$, become independent of $v$.
According to Eq. (\ref{niel}), the dependence of the denominator
on velocity is determined by the factor
\begin{eqnarray}
P'^*_{12}&+|G_1|^2/P'_2\approx
\Gamma_{ln}-i[\Omega_1-\Omega_2-({\bf
k}_1-{\bf k}_2){\bf v}]\nonumber\\
&+ [|G_1|^2/(\Gamma_{lg}+i\Omega_1)]-i(|G_1|/\Omega_1)^2{\bf
k}_1{\bf v}. \label{st}
\end{eqnarray}
Equation (\ref{st}) indicates a velocity-dependent broadening
($\Re\{|G_1|^2/P'_2\}$) and shift ($\Im\{|G_1|^2/P'_2\}$) of the
resonance that allows one to compensate for the Doppler effect at
$(|G_1|/\Omega_1)^2{\bf k}_1={\bf k}_1-{\bf k}_2$, and therefore
all molecules will be uniformly coupled to the two-photon
resonance independent of their velocities. Compensation of the
Doppler shifts with light shifts was proposed in \cite{Feok, Coh1,
Coh2}. Opportunities for great enhancement of various optical
processes with this effect were recently investigated in more
detail in \cite{{Vem},{4Sh},{4Ba},{EPJ},{Dong}}. We shall
illustrate the above considered effects with numerical simulations
applied to the experimental scheme described in Sec. \ref{pexp}.
\section{Numerical simulations}\label{ns}
As an example, we shall consider the transitions $l - g - n - m -
l$ (Fig. \ref{f1}) as those of sodium dimers Na$_2$:
$X'\Sigma_g^+(v"=0,J"=45)$ -- $A'\Sigma^+_u(6,45)(\lambda_1$= 655
nm) -- $X^1\Sigma_g^+(14,45) (\lambda_2$= 756  nm) --
$B^1\Pi_u(5,45) (\lambda_3$= 532 nm) -- $X'\Sigma_g^+(0,45)
(\lambda_4$= 480 nm) from the experiment \cite{3Hin}. We shall use
the experimental relaxation data:   $\gamma_{gl}$ = 7,
$\gamma_{gn}$ = 4, $\gamma_{mn}$ = 5, $\gamma_{ml}$ = 10,
$\Gamma_{l} = \Gamma_{n}$ = 20, $\Gamma_{g} = \Gamma_{m}$ = 120,
$\Gamma_{lm}=\Gamma_{nm}=\Gamma_{ng}=\Gamma_{lg}$ = 70,
$\Gamma_{ln}$ = 20, $\Gamma_{gm}$ = 120 (all in $10^{6}$ $s
^{-1}$). The Doppler width of the transition (FWHM) at the
wavelength $ \lambda _ 4 = 480 $  nm  at a temperature of
410$^\circ$C used for modelling is approximately equal to 1.7 GHz.
Then the Boltzmann population of level $ n $ is 1.4\% of that of
level $ l $. All of this, as well as the inhomogeneity of the
material parameters and phase mismatch $ \Delta k $, are accounted
for in the numerical simulations described below.
\subsection{Manipulating absorption and amplification indices with
coherent elimination of Doppler broadening of multiphoton
resonances}\label{aiedb} 
\begin{figure}[!h]
\includegraphics[height=.55\textwidth]{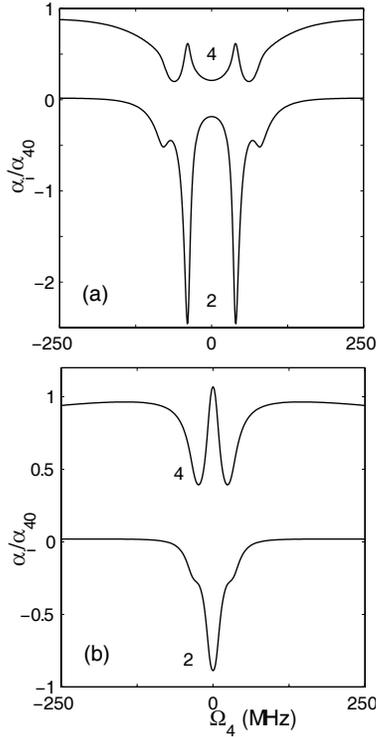}
\caption{\label{1bc}Spectral structures induced at low intensity
of the resonant driving fields. Upper plots - absorption of the
probe blue radiation, lower - Stokes gain.
$\Omega_1=\Omega_3=0$.\, (a) $G_{1}$ = 60 MHz, $G_{3}$ = 20 MHz;\,
(b) $G_1$ = 16 MHz, $G_3$ = 19 MHz. }
\end{figure}

\begin{figure}[!h]
\includegraphics[height=.45\textwidth]{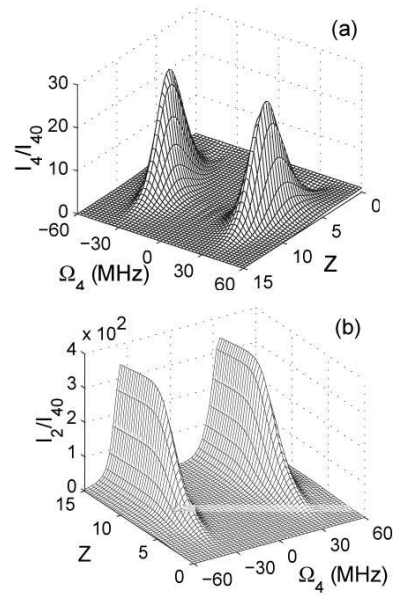}
\caption{\label{1efg} Laser-induced inversionless amplification,
transparency and generated Stokes radiation at low intensities of
fundamental beams ($Z=\alpha_{40}z$). $\Omega_1 = \Omega_3 = 0$,
$G_{10}$ = 60 MHz, $G_{30}$ = 20 MHz. \, (a) probe; (b) generated
Stokes idler radiation. }
\end{figure}

Figure \ref{1bc} depicts absorption
(upper plots) and Stokes amplification (lower plots) computed at
relatively low intensities of the driving fields that are set to
the corresponding resonances. Plot (a) corresponds to the
conditions at the the entrance to the medium, and  (b) to the
radiation depleted after propagation through the medium length
$Z=\alpha_{40}z$ = 15, where the OPA of the probe field reaches
its maximum (see  Fig. \ref{1efg}). As seen from the plots, the
depletion gives rise to a dramatic change in the spectral
properties of the absorption index for the probe blue field and of
the Stokes amplification index. The line shapes do not resemble
those for convention Raman processes. Despite some power-induced
depopulation of the lower level, an increase of $\alpha_4$ occurs
in some frequency interval in the vicinity of the transition
resonance center at the expense of its decrease in other frequency
intervals. For all the figures here and below, $\Omega_2$ is
determined by the equation $\Omega_2=\Omega_1+\Omega_3-\Omega_4$.
\begin{figure}[!h]
\includegraphics[height=.35\textwidth]{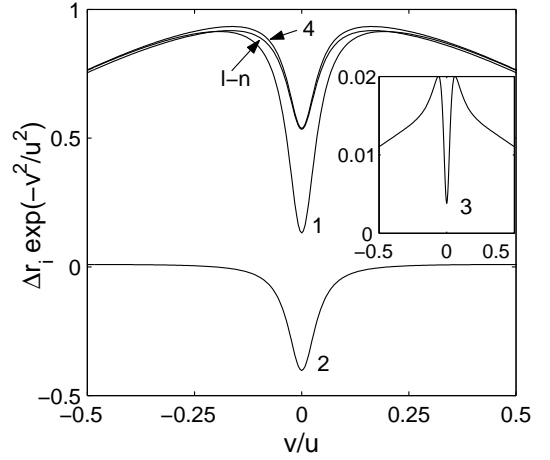}
\caption{\label{der} Differences of level population vs molecule
velocity projections v on the wave vectors ($\rm u$ is thermal
velocity). $l$-$n$ corresponds to $r_l-r_n$, and the other plots
to the transitions as indicated. All parameters are the same as in
Fig. \ref{1bc}(b). }
\end{figure}
\begin{figure}[!h]
\includegraphics[height=.6\textwidth]{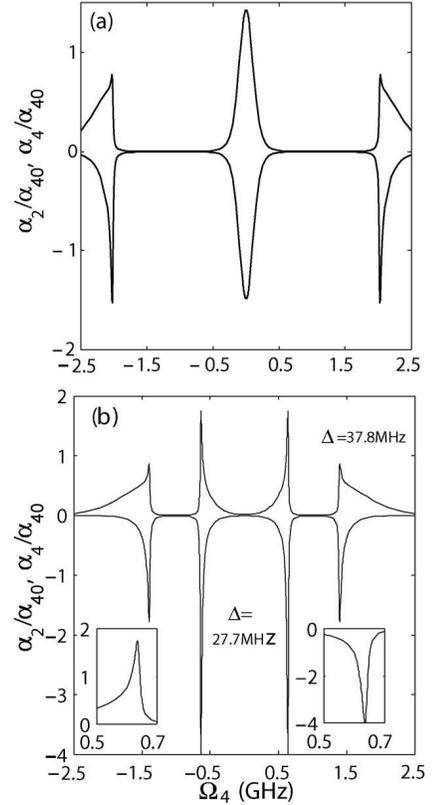}
\caption{\label{restruc}Resonance laser-induced spectral
structures (high intensities of driving fields). Upper plots -
absorption of the probe blue radiation, lower - Stokes gain.
$G_{1}$ = 1 GHz,\, $\Omega_3 = \Omega_1 = 0$.\, (a) $G_{3} = 1070$
MHz,\, (b)  $G_{3} = 415$ MHz. Insets - corresponding magnified
peaks. $\Delta$ - FWHM of the corresponding peaks.}
\end{figure}
%

\begin{figure}[!h]
\includegraphics[height=.35\textwidth]{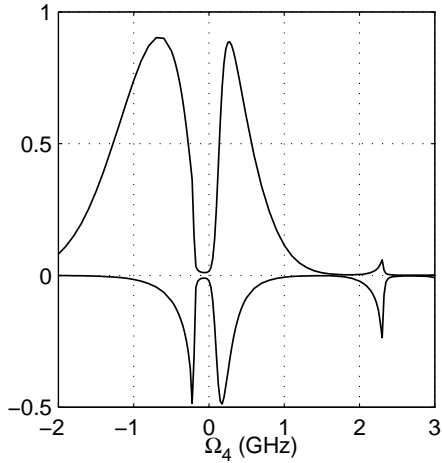}
\caption{\label{restruc1}Laser-induced spectral structures (high
intensities of driving fields). Upper plots - absorption of the
probe blue radiation, lower - Stokes gain. $G_{1}$ = 1 GHz,\,
$G_{3}$ = 400 MHz, \, $\Omega_1$ = 2140 MHz, $\Omega_3$ = 400 MHz.
}
\end{figure}
\begin{figure}[!h]
\includegraphics[height=.65\textwidth]{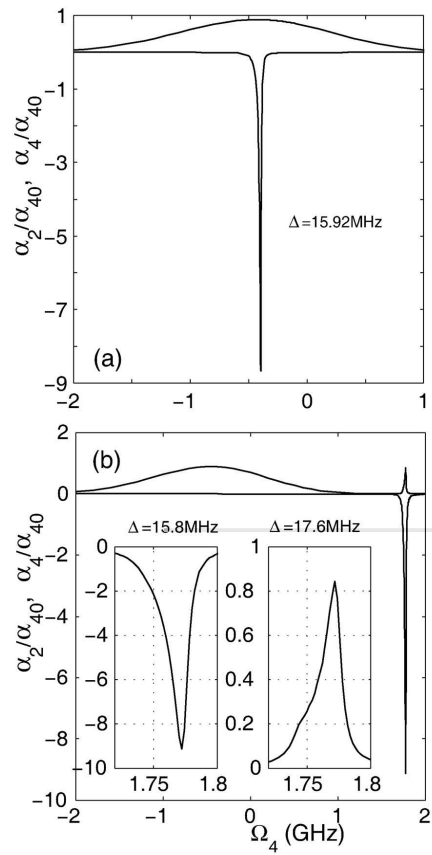}
\caption{\label{dc}Compensation of Doppler shifts with light
shifts and sub-Doppler resonances.  Upper plots - absorption of
the probe blue radiation, lower - Stokes gain. $\Omega_1$ = 2140
MHz, $G_{1}$ = 1 GHz. (a)\, $G_{3}$ = 0, ($\Omega_2$ = 2140 MHz -
$\Omega_4$);\, (b)\, $\Omega_3 = \Omega_1$, $G_{3} = 242$ MHz
($\Omega_2$ = 4280 MHz - $\Omega_4$). $\Delta$ is FWHM of the
corresponding peaks. Insets - magnified corresponding
laser-induced structures.\,  The Doppler FWHM of the Raman
transition is 170 MHz.}
\end{figure}
\begin{figure}[!h]
\includegraphics[height=.65\textwidth]{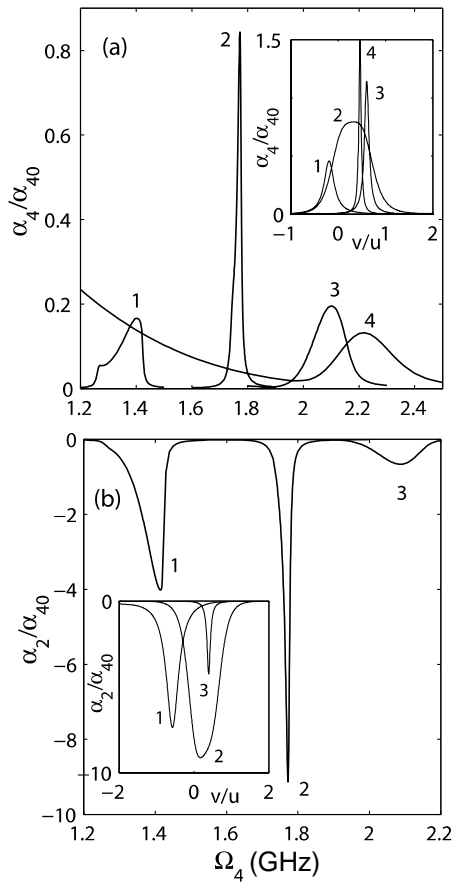}
\caption{\label{vd}Power-induced Doppler-free resonances.
$\Omega_1 = \Omega_3$ = 2140 MHz, $G_{3}$ = 242 MHz. \, (a) -
laser-induced structures in absorption index
$\alpha_4(\Omega_4)$,\, inset - velocity distribution
$\alpha_4(\rm v/\rm u)$ at $\Omega_4$ set to the center of the
corresponding resonance ($\rm u$ is thermal velocity). Maxwell's
envelope is removed. $\Delta$ - full width at half maximum of the
corresponding peaks.\, (b) - same for Stokes amplification
index.\, (a):
     1 --- $G_1$ = 1500 MHz ($\Delta$ = 98 MHz),
     2 --- $G_1$ = 1000 MHz ($\Delta$ = 17.6 MHz),
     3 --- $G_1$ = 500 MHz ($\Delta$ = 133 MHz),
     4 --- $G_1$ = 0 ($\Delta$ = 232 MHz);\,
(b) -- same parameters as in (a), where the FWHM of the structures
1, 2 and 3 are 64, 15.8 and 128 MHz. }
\end{figure}

Figure \ref{der} displays a substantial inhomogeneity of the
distribution of level populations over the velocities produced by
the driving fields. Population inversion appears at the Stokes
transition $ng$ in a narrow velocity interval, while no population
inversion is produced at the Raman transition $ln$. The figure
indicates the smallness of the fraction of molecules resonantly
coupled to the driving fields. This important fact must be taken
into account while considering amplification with and without
population inversion \cite{Kuch}.

Figure \ref{restruc}(a) shows corresponding line shapes at higher
intensities of the driving fields  chosen so that the splitting of
coupled levels is nearly equal. This determines the coincidence of
two out of the four spectral components at the probe and at the
Stokes transitions. Plot (b) demonstrates further substantial
change of the spectrum by variation of the intensity of one of the
fields and appearance of narrow sub-Doppler laser-induced
absorption and amplification structures. The position and shape of
the structures can be varied within the frequency-interval
comparable with the Doppler width.

Figure \ref{restruc1} illustrates the opportunities to form a
transparency window in any frequency interval in the vicinity of
the probe transition. As follows from Figs. \ref{1bc},
\ref{restruc} and \ref{restruc1}, the maxima and minimum in the
Stokes gain, as a rule, correspond to the maxima and the minimum
in the absorption index.

Figure \ref{dc} demonstrates the feasibilities of compensating for
Doppler shifts with ac-Stark shifts and corresponding substantial
field-induced narrowing of the induced resonances. The upper plot
in \ref{dc}(a) presents a red-shifted and slightly-decreased
Doppler-broadened absorption resonance, and the lower one the
feasibility of the formation of a power-induced Doppler-free
Stokes resonance. At $G_1^2\ll\Omega_1^2$, the ac-Stark shift of
the Stokes transition is estimated as $G^2/\Omega_1$, and the
dressed two-photon resonance as
$\Omega_2=\Omega_1+G_1^2/\Omega_1$. This is in a good agreement
with the computed plots. The FWHM of the induced resonance is 17
MHz, which is much less than the Doppler FWHM of the Raman
transition (170 MHz) and comparable to the homogeneous FWHM of the
Raman transition (6.4 MHz). Figure \ref{dc}(b) shows further
feasibilities for producing sub-Doppler structures with two
driving fields.

Figure \ref{vd} displays the sensitivity of the width of the
induced structures on the choice of the parameters of the driving
fields. The width of the most narrow resonance (plot 2) is even
less than the homogeneous FWHM of the corresponding transitions
and commensurate with that of the most narrow resonance in the
system (transition $ln$). Since all nonlinear resonances possess
asymmetric and non-Lorentzian shapes, the width of such a
resonance has different properties than those attributed to a
conventional resonance of a Lorentzian shape. The insets prove
that the narrowing is related to judicious compensation of the
Doppler shifts by ac-Stark shifts. Plots 2 in the insets indicate
that \textit{the central portion of the Maxwell velocity
distribution (the majority of atoms) becomes involved in the
coupling and simultaneously contributes to the absorption maximum
of the narrowed resonance} (compare with Fig. \ref{der}). The
other wider resonances remain inhomogeneously broadened.
\subsection{Generation of Stokes radiation, optical switching and
inversionless amplification of short-wavelength radiation above
the oscillation threshold in resonant optically-dense media
}\label{agwi}
\begin{figure}[!h]
\includegraphics[height=.6\textwidth]{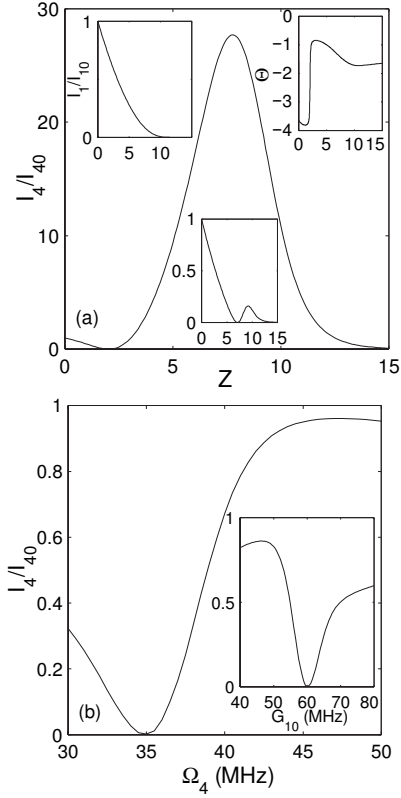}
\caption{\label{sw} Propagation of probe radiation (a) and optical
switching (b)\, ($Z=\alpha_{40}z$). $\Omega_1 = \Omega_3 = 0$,
$G_{10} = 60$ MHz, $G_{30} = 20$ MHz [(a) and main plot (b)]. (a)
$\Omega_4 = 35$ MHz (main plot) and  $\Omega_4 = 0$ (lower
 inset). Upper left inset - depletion of the fundamental radiation at $\omega_1$,
upper right inset - phase for main (a). \, (b) Main plot and inset
-  $Z = 2$, inset - $\Omega_4 = 35$ MHz. Minimum transmission is
less than $10^{-3}$.}
\end{figure}
\begin{figure}[!h]\vspace{-10pt}
\includegraphics[height=.45\textwidth]{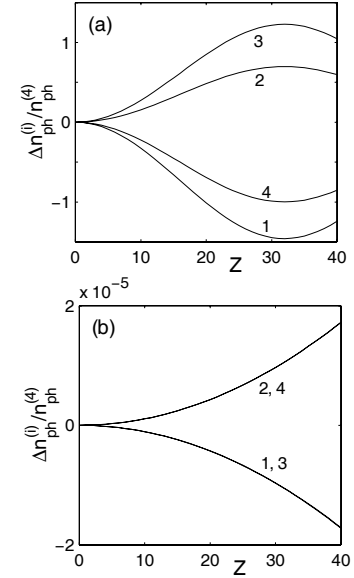}
\caption{\label{mr} Relative change of the photon numbers in the
coupled fields along the medium at neglected absorption and
refraction. $G_{10}$ = 60 MHz, $G_{30}$ = 20 MHz. (a)\, $\Omega_1
= \Omega_3 =\Omega_4= 0$, and
 (b)\,  $\Omega_1 = 1$ GHz, $\Omega_3 = -1$ GHz,  $\Omega_4=-2.5$ GHz.}
\end{figure}
\begin{figure}[!h]
\includegraphics[height=.5\textwidth]{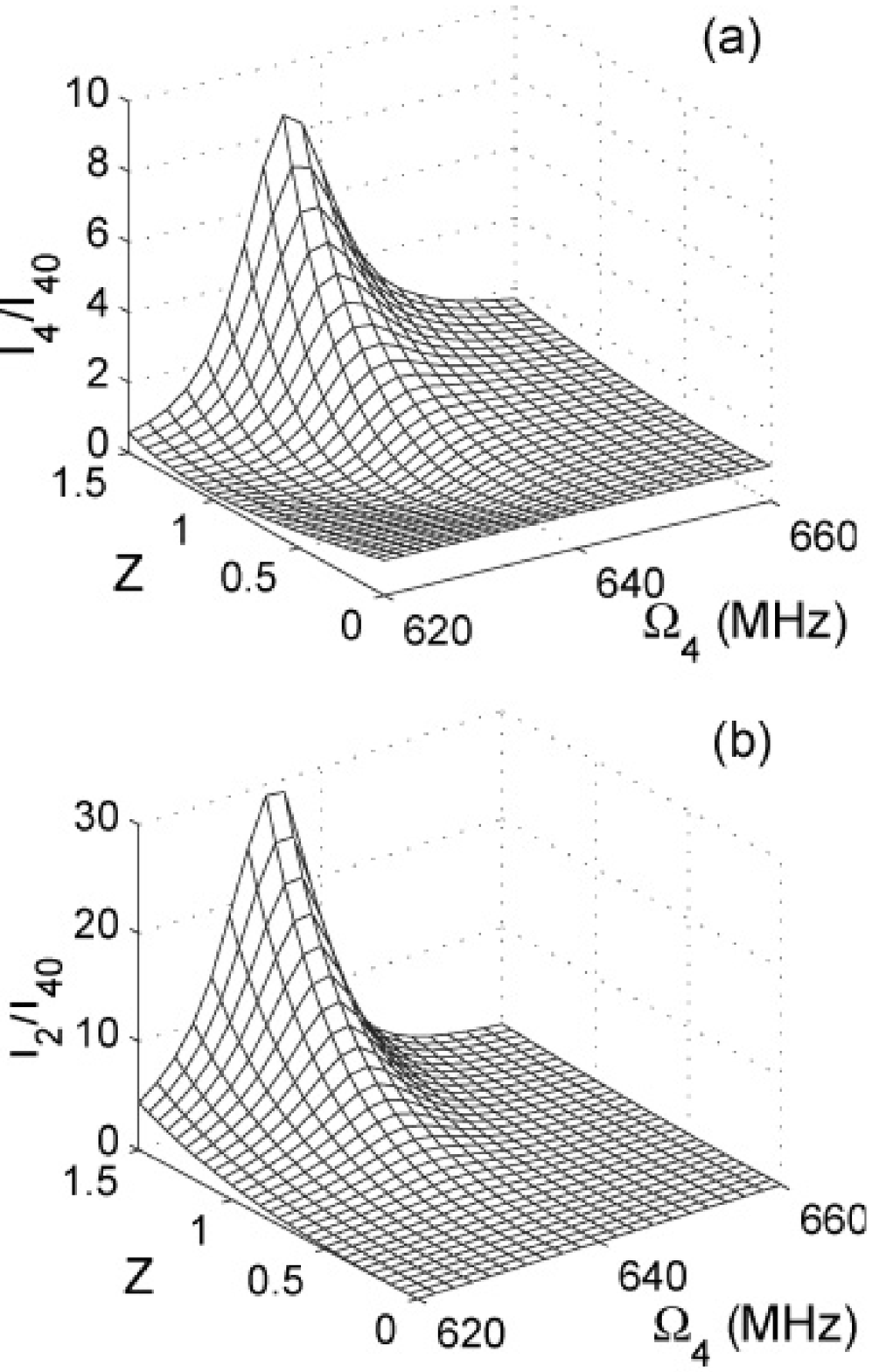}\\
\includegraphics[height=.5\textwidth]{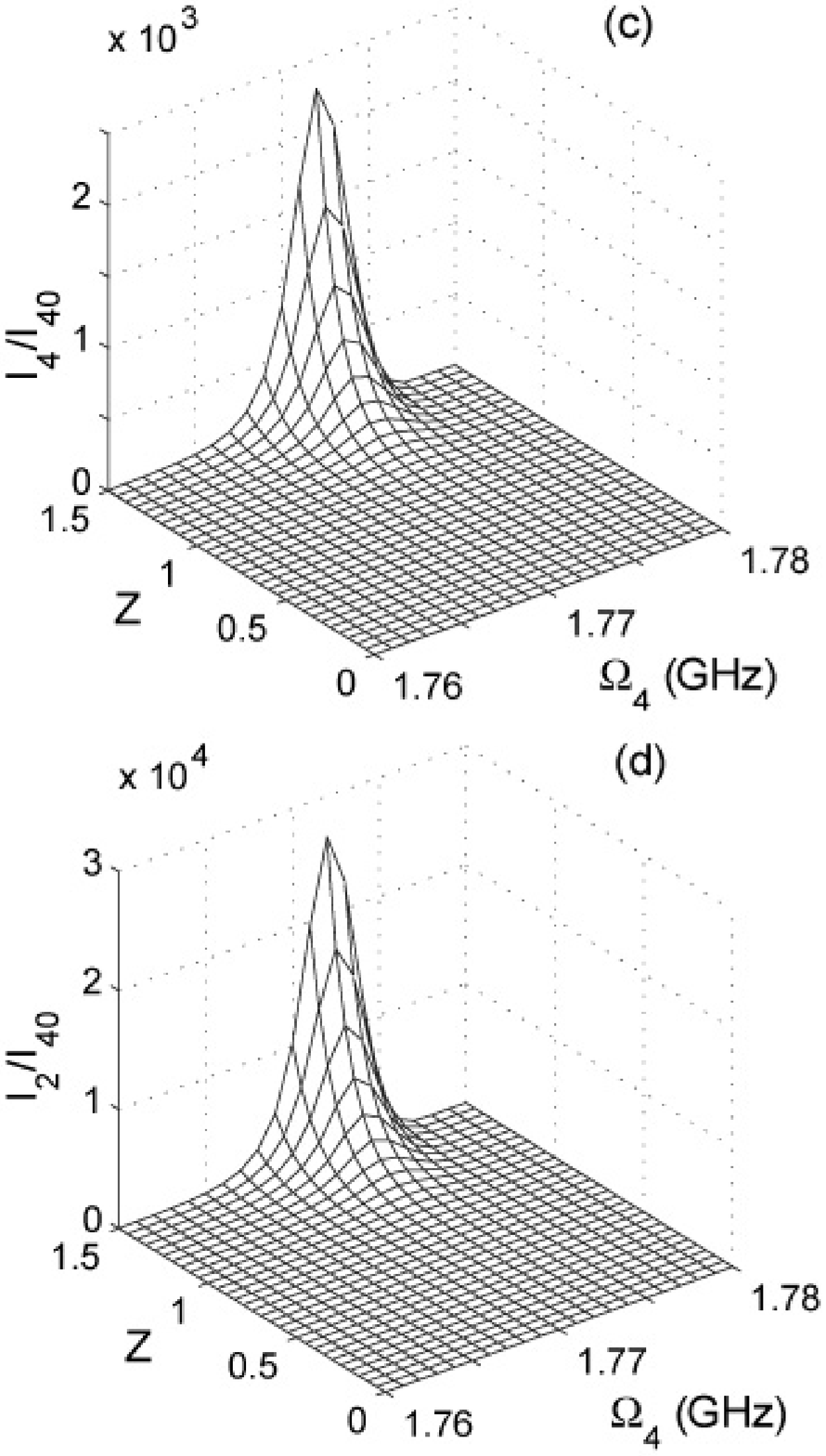}
\caption{\label{f34} Laser-induced inversionless amplification,
transparency and generated Stokes radiation at higher intensities
of fundamental beams.\, $G_{10}=1$ GHz. \, (a) and (b):
$\Omega_3=\Omega_1=0$,\, $G_{30}=415$ MHz.\, (c) and (d):
 $\Omega_3=\Omega_1=2140$, \,$G_{30}=242$ MHz.}
\end{figure}

Figure \ref{1efg} displays the results of numerical simulations of
amplification of short-wavelength radiation and of generation of
long-wavelength field controlled by quantum interference for the
conditions corresponding to Fig. \ref{1bc}. The optimum parameters
and output characteristics of the amplifier are stipulated from
the interplay between absorption, amplification of generated idler
Stokes radiation [Fig. \ref{1efg}(b)], phase-matching, FWM
cross-coupling parameters, and variation of all material
parameters along the medium due to saturation effects accounting
for the velocity-dependent NIE. It is remarkable that generated
idler Stokes radiation may several hundred times exceed the input
probe radiation [Fig. \ref{1efg}(b)]. The input Rabi frequencies
of the control fields for this graph correspond to the focused cw
radiations on the order of several tens of mW, i.e., to about one
photon per thousand molecules.  The length of the medium is scaled
to the absorption length at $\omega_4=\omega_{ml}$, with all
driving fields turned off.

In a good agreement with (\ref{opa}), Fig. \ref{sw}(a) shows that,
\textit{first, the probe beam is sharply depleted, and a
substantial optical length is required to achieve transparency,
and then there is significant inversionless gain} (main plot). The
complete transparency is achieved at a length of about 4.
\textit{With the same driving fields but with the probe radiation
tuned to the exact resonance} (lower right inset),
\textit{amplification is not achievable}. The upper inset shows
strong depletion of the driving radiation $E_1$ along the medium
at the chosen intensity and detuning from the resonance. The
evolution of the amplitudes of the fields along the medium is
determined by the relative phase $\Psi=\phi_{4}-\phi_{3}
+\phi_{2}-\phi_{1}+({\bf k}_4-{\bf k}_3+ {\bf k}_2-{\bf
k}_1)_zz=\Theta+\Delta k z$. The lower left inset in Fig.
\ref{sw}(a) shows substantial change of the relative phase
$\Theta$ along the medium that is determined by the inhomogeneity
of the coupled field and, consequently, material parameters.
Figure \ref{sw}(b) presents feasibilities of all-optical
switching, controlled by quantum interference. \textit{ Only a
small variation of the frequency of the probe field and/or the
input intensity of the driving field $I_{10}$ results in a
dramatic change of the intensity of the transmitted probe
radiation. By this, the sample can be easily driven to the opaque
state (less than $10^{-3}$ of the transparency) or to entire
transparency and strong amplification without any change of the
level populations.}

Figure \ref{mr}  displays the change of the number of photons in
each beam along the medium scaled to the number of photons in the
probe field at the entrance to the medium, assuming $\sigma_i=0$
in Eqs. (\ref{e42}) and (\ref{e13}). The plots are computed for
the same input intensities as Fig. \ref{sw}(a), but Fig.
\ref{mr}(a) corresponds to the fully-resonant case and Fig.
\ref{mr} (b) to far-from-resonance coupling. The figure proves
that \textit{the interference of quantum pathways plays such an
important role in fully-resonant multi-wavelength coupling that
the process as a whole can not be viewed as a sequence of
independent conventional elementary one-photon/two-photon FWM
processes, as they were introduced for off-resonant conditions in
the framework of perturbation theory. Indeed, Fig. \ref{mr}(a)
displays even qualitatively incorrect behavior with respect to the
Manley-Rowe relationship (photon conservation law). As detuning
from the resonance increases, the disagreement ceases}
[Figure~\ref{mr}(b)].

Figure \ref{f34} shows the feasibility of realizing \textit{strong
amplification of the probe short-wavelength radiation at the
expense of other longer-wavelength driving radiations, despite the
enhanced absorption index}. The importance of optimization of the
conversion parameters  based on the processes investigated here,
on the developed theory and on the numerical simulation is
explicitly seen from comparison of the plots (a) and (c). If so,
\textit{substantial (several thousand times) amplification of the
shortest wavelength radiation} becomes feasible.  Amplification
slightly exceeding unity ensures generation inside the ring
cavity. Hence, Figs. \ref{1efg}, \ref{sw} and \ref{f34} show
feasibilities of \emph{lasing without requirements of population
inversion at the resonant transition}.
\section{Earlier experiments on resonant nonlinear-optical
processes controlled through quantum interference}\label{rcqc}
\begin{figure}[!h]
\includegraphics[width=0.4\textwidth]{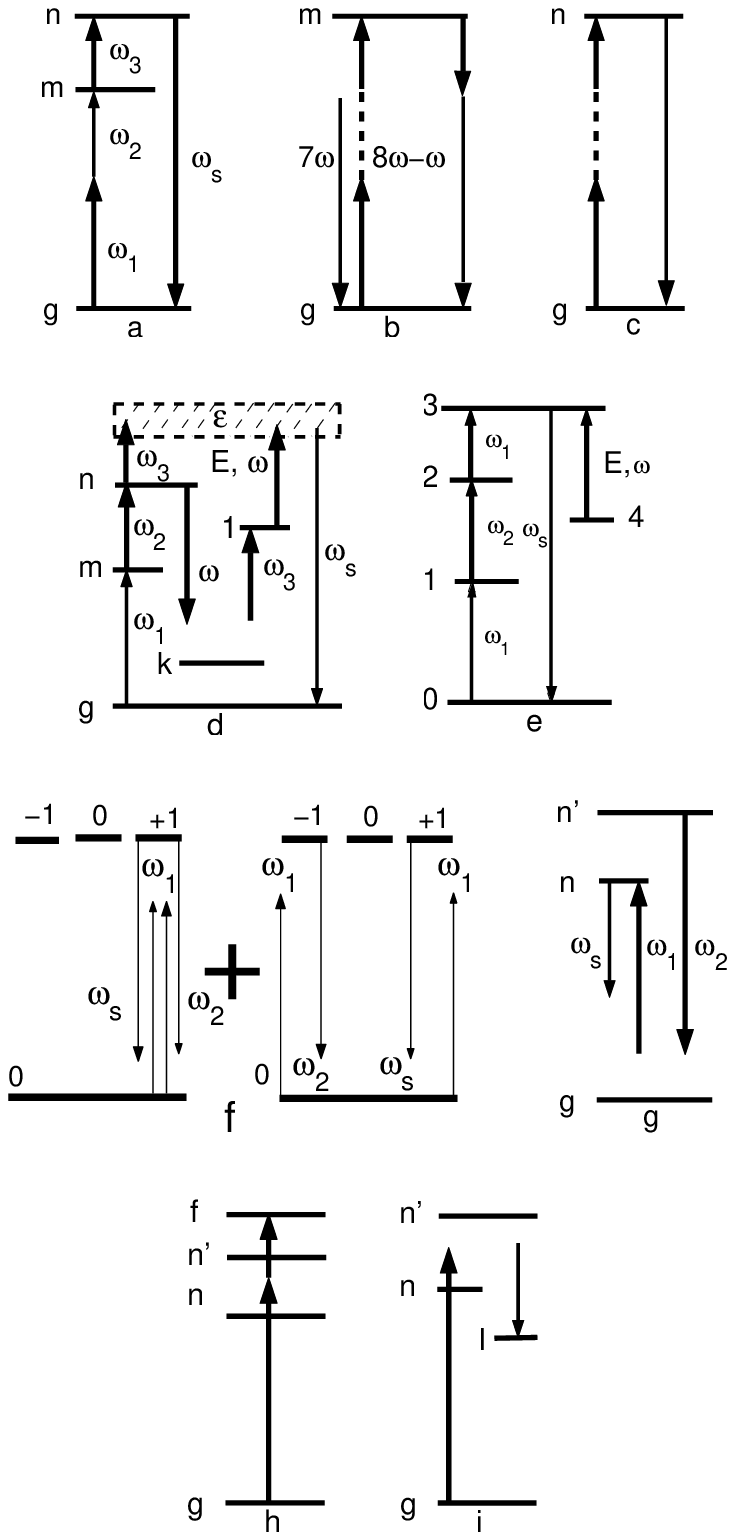}
\caption{\label{cqc} Interference of quantum pathways and quantum
control in resonant nonlinear optics.  (a) Up-conversion of
infrared radiation $\omega_2$; (b) interference of off-resonant
7th-order nonlinear polarization and 9th-order resonant
polarization at the generation of the 7th harmonic radiation; (c)
interference of resonant one-photon and multiphoton odd-order
nonlinear polarization; (d) laser-induced continuum structures
(LICS); (e) quantum control of sum-frequency mixing; (f) collision
induced four-wave mixing; (g) collision-induced resonance; (h)
interference of quantum pathways in two-photon absorption; and (i)
interference of quantum pathways in Raman scattering.}
\end{figure}
Constructive and destructive nonlinear interference effects, which
play a crucial role for quantum switching under consideration in
the present paper, are often met and play an important role in
nonlinear spectroscopy and in resonant nonlinear optics.
Comprehensive experiments on nonlinear interference effects at
Doppler-broadened transitions of Ne in the field of He-Ne lasers
had demonstrated feasibilities of producing transparency for cw
radiation at the transitions with unequal energy-level populations
and lasing without population inversion at the resonant transition
\cite{AwiB}. The examples of other early experiments where such
interference proved to play a critical role are depicted in Fig.
\ref{cqc}.

A weak infrared radiation field of arbitrary frequency $\omega_2$
[Fig. \ref{cqc}(a)] can be up-converted by a resonantly-enhanced
FWM process in the fields of two frequency-tunable  lasers,
$\omega_1$ and $\omega_3$. As confirmed in the experiments on Rb
vapors \cite{IR}, the process that fundamentally limits conversion
efficiency is destructive interference of two components in
coherence at transition $gm$ induced by $E_1$ and $E_2$ on one
hand, and by $E_3$ and $E_s$ on the other hand.

The interference of two nonlinear-optical processes in Hg vapor
[Fig. \ref{cqc}(b)] was employed to prove that direct conversion
of intense infrared radiation to VUV by an eight-photon-resonant
9th-order process is more efficient than that with a lower
seventh-order but off-resonant process \cite{Usp}. By a small
variation of the eight-photon detuning, destructive interference
can be turned to constructive. Thus higher-order resonant
polarization component was distinguished and compared with the
lower-order off-resonant one.

The interference between linear and nonlinear polarizations [Fig.
\ref{cqc}(c)] was shown to limit odd-order resonant harmonic
generation and multiphoton resonant ionization (see, e.g.
\cite{Usp} and references therein).

The interference of a one-photon transition to the dissociation or
ionization continuum and multiphoton transitions between discrete
states through the continuum stipulated by the same probe
radiation gives rise to laser-induced continuum structures
\cite{He,Kn} [Fig. \ref{cqc}(d)]. As was first shown in
\cite{GP2}, such laser-induced autoionizing-like  structures in
the continuum allow one the increase FWM polarization while
decrease the absorption of generated short-wavelength radiation.
Laser-induced continuum structures enable quantum control of
chemical processes related to dissociation and ionization by
employing constructive and destructive interference. (For a recent
literature survey of the experiments, see \cite{PKG04}). Similar
processes can be realized in discrete transitions [Fig.
\ref{cqc}(e)].

Figure \ref{cqc}(f) illustrates the FWM process
$2\omega_1=\omega_s+\omega_2$, where $E_1$ and $E_2$ are linearly
and orthogonal polarized waves with nearly equal frequencies close
to the frequency of the transition between non-degenerate level
and degenerate level which consist of three Zeeman sub-levels.
Linear polarized oscillations can be treated as combinations of
two oppositely polarized circular oscillations. Each
circularly-polarized component of $E_s$ can be created both due to
processes involving photons of the same  polarization [left side
of Fig. \ref{cqc}(f)] and processes involving photons of opposite
circular polarization [right side of Fig. \ref{cqc}(f)]. As was
proved in experiments with transitions between excited levels of
Ne \cite{CI}, destructive interference of these two channels
eliminate FWM process completely unless specific collisions occur
or even very weak magnetic or electric fields are applied. This
allows one to selectively detect such collisions among many other
relaxation processes as well as to measure the weak fields.

In a similar way, the FWM process presented in Fig. \ref{cqc}(g)
does not feel the resonance at $\omega_2-\omega_1=\omega_{n'n}$
between excited unpopulated levels unless collisions disbalance
the destructive interference in the quantum system \cite{Bl1,Bl2}.
Figures \ref{cqc}(i) and (k) show possible destructive
interference of two quantum pathways through intermediate levels
$n$ and $n'$ that allows one to control the processes of
two-photon excitation.

\section{Conclusions}\label{c}
Specific features of manipulating linear and nonlinear optical
properties of resonant materials through constructive and
destructive quantum interference in three- and four-level  media
with \emph{inhomogeneously-broadened multiphoton transitions} are
studied. The feasibility of compensating for Doppler-shifts with
power-shifts by means of quantum control is shown. This enables
the removal of a fundamental limitation on resonant multiphoton
coupling with molecules from wide velocity-intervals concurrently
independent of Doppler shifts of their resonances and, thus,
enables significant increase of the cross-sections of various
nonlinear-optical processes in warm gases. Similar effects, with
the exclusion of the angular anisotropy of laser-induced nonlinear
resonances, are achievable at inhomogeneously-broadened
transitions in solids.

The propagation of a probe signal in an extended initially
strongly-absorbing resonant sample is investigated, which is
controlled by two driving fields through the interference of
resonant Raman-type and four-wave mixing processes. \emph{It is
shown that under fully- and near-resonant coupling, the
interference of quantum pathways usually plays so important a role
that the propagation processes as a whole cannot be viewed as a
sequence of independent conventional single-photon, multiphoton
and FWM processes, as they can be introduced for off-resonant
coupling in the framework of perturbation theory.} Substantial
differences from the behavior that would be expected based on the
concept of linear and nonlinear off-resonance processes is
demonstrated through numerical simulations. As regarding the
proposed scheme, \emph{it is shown that the entire process cannot
be viewed as resonant optical parametric amplification accompanied
by Stokes gain and absorption. However, it acquires such features
with an increase of resonance detunings.}

Applications to photon switching, frequency-tunable narrow-band
filtering, and amplification without population inversion of the
probe beam are explored through quantum control by two
longer-wavelength fields. Switching to enhanced (practically
absolute) opacity or to perfect transparency, or even to
amplification without population inversion above the oscillation
threshold, is shown to be possible due to quantum interference and
to readily-achievable Stokes gain. Such gain appears to be
inherent to the far-from-degenerate double-Lambda scheme under
consideration. The \emph{feasibilities of manipulating Raman gain
for generated idler infrared radiation} are explored too. The
outlined processes are associated with the appearance of the
complex narrow laser-induced spectral structures in absorption,
refraction, Raman,  and FWM susceptibilities that vary along the
medium. As a rule, a decrease in the absorption index is
accompanied by a decrease in local nonlinear-optical coupling, and
vice versa. However, the outlined opportunities of strong
amplification becomes possible through the deliberate trade-off
optimization.

It is shown that \emph{for fully-resonant fundamental and probe
fields, which would correspond to Raman resonance for the
generated idler radiation, the probe signal is only depleted}.
Because of destructive interference, the rate of depletion may
become even higher than that in the absence of the driving fields.
On the other hand, \emph{there exist detunings where the initial
depletion changes for amplification after propagation through a
substantial medium length}. It is shown that one can judiciously
select parameters such that the transition from absorption to
amplification through complete transparency becomes very sharp.
This allows one to vary the transparency of the probe radiation or
to switch its transparency between components of wavelength
division multiplexing at optical networking by a small variation
of intensity and/or detuning of the control fields from
resonances. It is shown that the \emph{control can be achieved
with the intensities of the driving fields at the level of
conventional cw lasers} of about one photon per one molecule. The
feasibilities of producing amplification well above the
oscillation threshold and of \emph{lasing without population
inversion at short-wavelength transitions} at the expense of
lower-frequency control fields are demonstrated.

The outcomes are illustrated with numerical experiments. Although
the data relevant to most typical and easy-to-realize experiments
with sodium molecules have been employed for numerical
simulations, similar energy level configurations with
inhomogeneously broadened transitions are easily found in
rare-earth-doped solids used as Raman amplifiers in optical
networking and in quantum nanostructures. Proof-of-principle
experiments on fast optical switching based on quantum control and
FWM have proven such feasibilities \cite{Sw}. Since the principles
of the proposed switching are based on ground-level coherence, the
implementation of spin coherence with the lowest relaxation rate
allows one to improve dramatically the performance of such quantum
switches.
\begin{acknowledgments}
This work was supported in part by DARPA under grant No. MDA
972-03-1-0020. A.~K.~P. thanks Vladimir Shalaev for discussions of
possible applications of similar schemes of quantum control in
nanophotonics.
\end{acknowledgments}
\appendix\section{Nonlinear susceptibilities and energy level  populations for the
case where each level is coupled to only one driving field: open
and closed schemes}\label{susc}
 With the aid of the solution of a set of equations (\ref{od}) for the
off-diagonal elements of the density matrix up to first order in
perturbation theory with respect to the weak fields, the equations
for the susceptibilities can be presented as \cite{{1PSPI},{MysK}}
\begin{eqnarray}
\dfrac{\chi_{i}}{\chi_{i}^0}= \dfrac{\Gamma_{i}{\Delta r_{i}}}{{P_i}{\Delta n_i}} (i=1,3),
\dfrac{\chi_{i}}{\chi_{i}^0}=\dfrac{\Gamma_{i}{R_{i}}}{{P_i}{\Delta n_i}} (i=2,4),\label{ch1}&&\\
\tilde\chi_2 =-iN\dfrac{d_{ml}d_{lg}d_{gn}d_{nm}/8\hbar^3}{d_2(1+v_5^*+g_5^*)}\nonumber&&\\
\times\left[\left(\dfrac{\Delta r_1}{P_1P_{41}^*}+ \dfrac{\Delta
r_3}{P_3P_{43}^*}\right)+ \dfrac {R_4^*}{P_4^*}\left(\dfrac
1{P_{41}^*}+\dfrac 1{P_{43}^*}\right)\right],\label{ch2}&&\\
\tilde\chi_4 = -iN\dfrac{d_{ml}d_{lg}d_{gn}d_{nm}/8\hbar^3}{d_4(1+v_7^*+g_7^*)}\nonumber&&\\
\times\left[\left(\dfrac{\Delta r_1}{P_1P_{12}}+ \dfrac{\Delta
r_3}{P_3P_{32}}\right)+ \dfrac {R_2^*}{P_2^*}\left(\dfrac
1{P_{12}}+\dfrac 1{P_{32}}\right)\right].\label{ch3}&&
\end{eqnarray}\\

Here $\chi_{i}^0=\chi_{i}(G_i=0, \Omega_i=0)$ is a resonance value
of the susceptibility for all fields turned off,
\begin{eqnarray*}
g_1&=&{|G_1|^2}/{P_{41}P_1^*}, g_2={|G_1|^2}/{P_{12}^*P_2},
g_3={|G_1|^2}/{P_{12}^*P_1^*},\\
g_4&=&{|G_1|^2}/{P_{41}P_4},
g_5={|G_1|^2}/{P_{43}d_2^*}, g_6={|G_1|^2}{P_{41}d_2^*},\\
g_7&=&{|G_1|^2}/{P_{32}^*d_4^*}, g_8={|G_1|^2}/{P_{12}^*d_4^*},
v_1={|G_3|^2}/{P_{43}P_3^*},\\
v_2&=&{|G_3|^2}/{P_{32}^*P_2},
v_3={|G_3|^2}/{P_{32}^*P_3^*}, v_4= {|G_3|^2}/{P_{43}P_4},\\
v_5&=& {|G_3|^2}/{P_{41}d_2^*}, v_6= {|G_3|^2}/{P_{43}d_2^*},
v_7={|G_3|^2}/{P_{12}^*d_4^*},\\
v_8&=&{|G_3|^2}/{P_{32}^*d_4^*},\end{eqnarray*}
\begin{widetext}
\begin{eqnarray}
R_2&=&\dfrac { \Delta r_2(1+g_7+v_7)-v_3(1+v_7-g_8)\Delta
r_3-g_3(1+g_7-v_8)\Delta r_1}
{(1+g_2+v_2)+[g_7+g_2(g_7-v_8)+v_7+v_2(v_7-g_8)]},
\label{R2}\\
R_4&=&\dfrac { \Delta r_4(1+v_5+g_5)-g_1(1+g_5-v_6)\Delta
r_1-v_1(1+v_5-g_6)\Delta r_3}
{(1+g_4+v_4)+[v_5+v_4(v_5-g_6)+g_5+g_4(g_5-v_6)]}.\label{R4}
\end{eqnarray}
\end{widetext}
The populations are described by the formulas below.\\

\noindent OPEN CONFIGURATION:
\begin{eqnarray}\label{eq5}
\Delta r_1&=&\frac{{(1+\ka_3)\Delta n_1+b_1\ka_3 \Delta n_3}}{
{(1+\ka_1)(1+\ka_3)-a_1\ka_1b_1\ka_3}},\nonumber\\
\Delta r_3&=&\frac{{(1+\ka_1)\Delta n_3+a_1\ka_1 \Delta n_1}}{
{(1+\ka_1)(1+\ka_3)-a_1\ka_1b_1\ka_3}},\\
\Delta r_2&=&\Delta n_2-b_2\ka_3\Delta r_3-a_2\ka_1\Delta r_1,\nonumber\\
\Delta r_4&=&\Delta n_4-a_3\ka_1\Delta r_1-b_3\ka_3\Delta
r_3,\nonumber
\end{eqnarray}
\begin{eqnarray}
r_m&=&n_m+(1-b_2)\ka_3\Delta r_3,\nonumber\\
r_g&=&n_g +(1-a_3)\ka_1\Delta
r_1,\nonumber\\
r_n&=&n_n-b_2\ka_3\Delta r_3+a_1\ka_1\Delta r_1,\\
r_l&=&n_l-b_1\ka_3\Delta r_3+a_3\ka_1\Delta r_1,\nonumber
\end{eqnarray}
where
\begin{eqnarray*}
\ka_1&=&\ka_1^0{\Gamma_{lg}^2}/{|P_1|^2},\quad
\ka_3=\ka_3^0{\Gamma_{mn}^2}/{|P_3|^2},\nonumber\\
\ka_1^0&=&{2(\Gamma_l+\Gamma_g-\gamma_{gl})}|G_1|^2/
{\Gamma_l\Gamma_g\Gamma_{lg}},\nonumber\\
\ka_3^0&=&{2(\Gamma_m+\Gamma_n-\gamma_{mn})}|G_3|^2/
{\Gamma_m\Gamma_n\Gamma_{mn}},
\end{eqnarray*}
\begin{eqnarray*}
a_1=\dfrac{\gamma_{gn}a_2}{\Gamma_n-\gamma_{gn}}=
\dfrac{\gamma_{gn}\Gamma_la_3}{\Gamma_n(\Gamma_g-\gamma_{gl})}=
\dfrac{\gamma_{gn}\Gamma_l}{\Gamma_n(\Gamma_l+\Gamma_g-\gamma_{gl})},&&\\
b_1=\dfrac{\gamma_{ml}\Gamma_nb_2}{\Gamma_l(\Gamma_m-\gamma_{mn})}
=\dfrac{\gamma_{ml}b_3}{\Gamma_l(\Gamma_l-\gamma_{ml})}
=\dfrac{\gamma_{ml}\Gamma_n}{\Gamma_l(\Gamma_m+\Gamma_n-\gamma_{mn})}.&&
\end{eqnarray*}\\

\noindent CLOSED CONFIGURATION:\\
\\
In this case, the populations of levels are given by the equations
\begin{eqnarray}\label{syst}
\Gamma_mr_m&=&w_mr_l-2\Re\left\{iG_3^*r_3 \right\},\nonumber \\
\Gamma_gr_g&=&w_gr_l-2\Re\left\{iG_1^*r_1 \right\},\nonumber \\
\Gamma_nr_n&=&w_nr_l+2\Re\left\{iG_3^*r_3 \right\}\\
&+&\gamma_{gn}r_g+\gamma_{mn}r_m,\nonumber \\
r_l&=&1-r_m-r_g-r_n,\nonumber
\end{eqnarray}\pagebreak
whose solution is
\begin{align}\label{solve}
&r_l=n_l(1+\ka_3)(1+\ka_1)/\beta,\nonumber \\
&r_g=(1+\ka_3)[n_l(1+\ka_1)-\Delta n_1]/\beta,\nonumber  \\
&r_n=\left\{n_m(1+\ka_3)(1+\ka_1)\right.\\
&+[\Delta n_3(1+\ka_1)
 +\Delta n_1\gamma_2\ka_1/\Gamma_n](1+b\ka_3)\left.\right\}/\beta,\nonumber \\
&r_m=\left\{n_m(1+\ka_3)(1+\ka_1)\right.\nonumber \\
&+[\Delta n_3(1+\ka_1) +\Delta n_1\gamma_2\ka_1/\Gamma_n]
b\ka_3\left.\right\}/\beta,\nonumber  \end{align}
\begin{align}\label{dr}
&\Delta r_3=r_n-r_m=\left[\Delta n_3(1+\ka_1) +\Delta
n_1\gamma_2\ka_1/\Gamma_n\right]/\beta,\nonumber \\
&\Delta r_1=r_l-r_g=\Delta n_1(1+\ka_3)/\beta.
\end{align}
Here,
\begin{eqnarray*}
\Delta n_1&=&n_l-n_g,
\Delta n_3=n_n-n_m,\nonumber\\
n_m&=&n_lw_m/\Gamma_m, n_g=n_lw_g/\Gamma_g,
n_n=n_l{w_n}'/\Gamma_n, \nonumber\\
n_l&=&(1+
w_m/\Gamma_m+w_g/\Gamma_g+{w_n}'/\Gamma_n)^{-1},\nonumber\\
{w_n}'&=&w_n+w_g\gamma_{gn}/\Gamma_n+w_m\gamma_{mn}/\Gamma_n,\nonumber\\
b&=&\Gamma_n/(\Gamma_m+\Gamma_n-\gamma_3).\nonumber
\end{eqnarray*}
\begin{eqnarray*}
\ka_1&=&\dfrac{({2|G_1|^2}}{{\Gamma_1\Gamma_g})({\Gamma_1^2}/{|P_1|^2})},
\ka_3=\dfrac{{2|G_3|^2(\Gamma_m+\Gamma_n-\gamma_3)}}{{\Gamma_m\Gamma_n\Gamma_3}(
{\Gamma_3^2}/{|P_3|^2})},\nonumber\\
\beta&=&(1+\ka_3)[1-\Delta n_3+2(n_l+n_m)\ka_1]\nonumber\\
&+&(1+2b\ka_3)[\Delta n_3(1+\ka_1)+\Delta
n_1\gamma_2\ka_1/\Gamma_n].\nonumber \end{eqnarray*} The remaining
notations are as before.

\end{document}